\documentclass[aps,pre,twocolumn,superscriptaddress,showpacs]{revtex4-1}
\usepackage{graphicx}
\usepackage{dcolumn}
\usepackage{bm}
\usepackage{epsfig}
\usepackage{color}

\begin{document}


\title{A Dynamical Trap Made of Target-Tracking Chasers}


\author{Guo-Jie Jason Gao}
\email{koh.kokketsu@shizuoka.ac.jp, gjjgao@gmail.com}
\affiliation{Department of Mathematical and Systems Engineering,
  Shizuoka University, Hamamatsu, Shizuoka 432-8561, Japan}

\date{\today}

\begin{abstract}
We propose a dynamical trapping system composed of multiple chasers
subject to target-tracking forces utilizing the velocity and position
information of a single escaping target. To successfully capture the
target, dividing chasers into multiple groups while each group
approaching its assigned destination in the proper vicinity of the
target is essential. Moving direction synchronization between the
target and its chasers is also crucial to the capturing process, while
guiding chasers to the predicted position of the target in future only
improves the efficiency of capture but is not indispensable. Potential
applications of our trapping system include capturing live animals
such as bears invading a human residential area.
\end{abstract}


\maketitle

\section{Introduction}
\label{introduction}
Wildlife-human conflict increases significantly due to climate change
\cite{abrahms23}. For example, warmed temperature prolongs bears'
active season and therefore they have to look for additional food
resources in humans' residential areas, which ensues greater
human-caused mortality \cite{laufenberg18}. In Japan, a similar
temperature issue has caused the bear crisis in these years
\cite{honda20,nakamoto25}, and human mortality reached a historical
high in the year of 2025 \cite{moej25}. Although there exist long-term
bear forecasting and warning systems \cite{oka04,koike09,nakamoto25}
and quick but risky solutions such as using a hunting rifle, pepper
spray, or bear bell for wildlife management, but an effective and
non-lethal strategy to neutralize imminent danger from individual
bears invading areas inhabited by humans without exposing either
creature getting involved to unpredictable consequence is still
lacking. In this study, we fulfill this need by proposing a dynamical
trap composed of multiple chasers based on a prescribed algorithm that
automatically tracks and confines a bear (target) - a kind of
interaction also commonly found during the early stage when predators
hunt preys in the biological world.

We take reference to the animal collaborative hunting strategies and
patterns \cite{hansen23} and the existing agent-based numerical
target-chaser models
\cite{kamimura10,masuko17,janosov17,zhang19,bernardi22,souza22,su23}. Then
we add new functions that can be performed easily by artificial
machines but not animals such as maneuvering by accurately monitoring
the relative distance to a target. Specifically, the target escapes if
a chaser approaches too close, while all chasers constantly follows
the target by tracking its position and velocity. We simulate their
interactions using the discrete element method (DEM), where each
individual (target or chaser) is subject to modeled inter-individual
forces and obeys Newton's equation of motion, integrated numerically
as a function of time.

Below we elaborate on the details of the simulated system in section
\ref{method}, followed by quantitative analysis in section
\ref{results_and_discussions}. We conclude our study in section
\ref{conclusions}.

\section{Numerical simulation method}
\label{method}
In the DEM simulation, we place $N$ individuals composed of one target
and $N-1$ chasers on a two-dimensional $x$-$y$ plane with open
boundary conditions. We choose a two-dimensional setup because bears
mostly move on the ground with ignorable altitude variation on a short
time scale. The target, randomly surrounded by static chasers at the
beginning, has a nonzero initial velocity. Each individual (target or
chaser) $i$ with mass $m$ and subject to total force
${{\mathord{\buildrel{\lower3pt\hbox{$\scriptscriptstyle\rightharpoonup$}}
      \over F} }_i}$ is treated as a sizeless particle and accelerates
by
${{\mathord{\buildrel{\lower3pt\hbox{$\scriptscriptstyle\rightharpoonup$}}
      \over a} }_i}$, following Newton's translational equation of
motion
\begin{equation} \label{newton_law}
m\mathord{\buildrel{\lower3pt\hbox{$\scriptscriptstyle\rightharpoonup$}}
  \over a} _i =
\mathord{\buildrel{\lower3pt\hbox{$\scriptscriptstyle\rightharpoonup$}}
  \over F} _i =
\mathord{\buildrel{\lower3pt\hbox{$\scriptscriptstyle\rightharpoonup$}}
  \over F} _i^{rep} +
\mathord{\buildrel{\lower3pt\hbox{$\scriptscriptstyle\rightharpoonup$}}
  \over F} _i^{self} +
c\mathord{\buildrel{\lower3pt\hbox{$\scriptscriptstyle\rightharpoonup$}}
  \over F} _i^{track},
\end{equation}
where
$\mathord{\buildrel{\lower3pt\hbox{$\scriptscriptstyle\rightharpoonup$}}
  \over F} _i^{rep}$,
$\mathord{\buildrel{\lower3pt\hbox{$\scriptscriptstyle\rightharpoonup$}}
  \over F} _i^{self}$, and
$\mathord{\buildrel{\lower3pt\hbox{$\scriptscriptstyle\rightharpoonup$}}
  \over F} _i^{track}$ are repulsive force from other interacting
particles, self-regulation force, and tracking force,
respectively. The parameter $c$ is a constant whose value is $0$ for
the target and $1$ for all chasers.  We integrate
Eqn. \ref{newton_law} using the velocity Verlet algorithm
\cite{tildesley17}. Below we give details of these constituent forces.

To simulate a group of separate particles, we consider the nonlinear
elastic repulsive force that sets a lower limit to the interparticle
distance \cite{charbonneau17}. The repulsive force on particle $i$
having $n$ interacting neighbors $j$ can be expressed as
\begin{equation} \label{interparticle_rep_force}
\mathord{\buildrel{\lower3pt\hbox{$\scriptscriptstyle\rightharpoonup$}}
  \over F} _i^{rep} = \sum\limits_{j \ne i}^n \epsilon {\left(
  {\frac{{{\delta _{ij}}}}{{{d_{ij}}}}} \right)^{3/2}}\Theta ({\delta
  _{ij}}){{\hat n}_{ij}},
\end{equation}
where $\epsilon$ is the elastic repulsive force amplitude, $d_{ij} =
(d_i+d_j)/2$ is the average interaction diameter of particles $i$ and
$j$, $\delta_{ij} = d_{ij} - r_{ij}$ is the repulsive interaction
overlap, $r_{ij}$ is the center-to-center distance between particles
$i$ and $j$, $\Theta(x)$ is the Heaviside step function, and ${{\hat
    n}_{ij}}$ is the unit vector pointing from the center of particle
$j$ to that of particle $i$. Practically, free-roaming animals such as
bears show mild behavioral response to approaching drones
\cite{ditmer15}. Therefore, we use the smooth repulsive force defined
in Eqn. \ref{interparticle_rep_force} to mimic the escaping behavior
of a bear (target), where the target feels an approaching chaser and
runs away from it. Throughout the study, we set $\epsilon^t = 25$ (for
target-chaser interaction) and $d^t = 0.2$ for the target, and
$\epsilon^c = 20$ (for chaser-chaser interaction) and $d^c = 0.05$ for
chasers.

The self-regulation force of particle $i$ contains a braking term and
a random term
\begin{equation} \label{self_reg_force}
\mathord{\buildrel{\lower3pt\hbox{$\scriptscriptstyle\rightharpoonup$}}
  \over F} _i^{self} = \mu ({v_0} - {v_i}){{\hat v}_i} +
R^{(x,y)}_{\eta} {{\hat n}^{(x,y)}},
\end{equation}
where in the braking term $v_0$ is the goal braking velocity, $v_i$ is
its current velocity along its unit vector ${\hat v}_i$, and $\mu$
sets the magnitude of the braking force. If a particle does not chase
or escape from others, it stays still.  Therefore we set $v_0=0$ to
model this scenario. In the random term, $R^{(x,y)}_{\eta}{\hat
  n}^{(x,y)}=R^x_{\eta}\hat x + R^y_{\eta}\hat y$, where both
$R^x_{\eta}$ and $R^y_{\eta}$ are an independent random number uniform
between [$-R_{\eta}$,$R_{\eta}$] and $\hat x$ and $\hat y$ are unit
vectors in the $x$ and $y$ directions, respectively. In the study, we
set $\mu=10$ and $R_{\eta}=0.1$.

\begin{figure}
\includegraphics[width=0.40\textwidth]{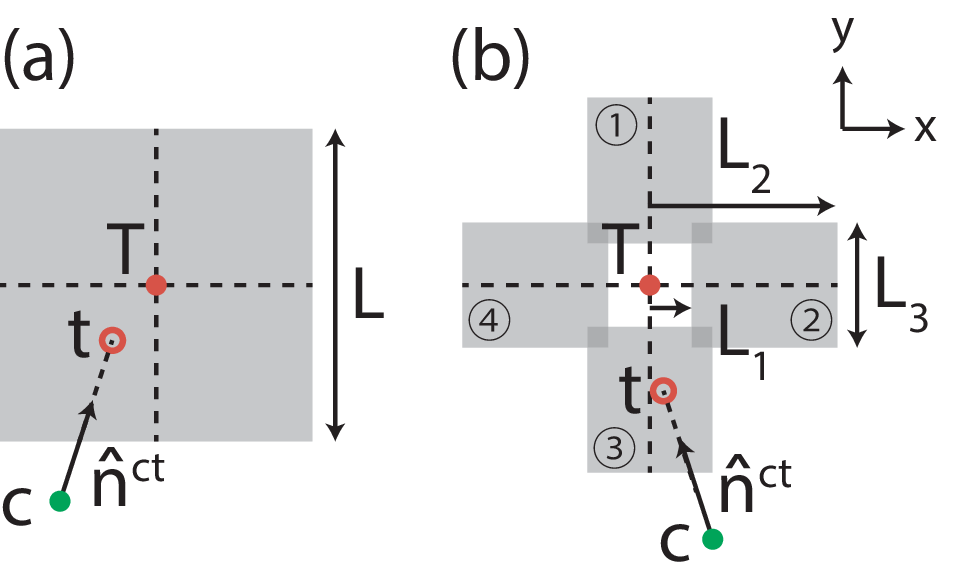}
\caption{\label{fig:algorithm} (Color online) Specifications of the
  position-related term $\beta \hat n^{ct}$ of the tracking force. A
  chaser pursuing the target is subject to a force of magnitude
  $\beta$ along the center-to-center direction $\hat n^{ct}$ between
  the chaser's current position $c$ (filled green circle) and a random
  point $t$ (open red circle) in (a) a square domain of side length
  $L$ or (b) a cross-like domain, with the shortest distance $L_1$
  away from the target and composed of four identical sub-domains of
  size $L_2-L_1$ by $L_3$, centered on $T$ (filled red circle), the
  target's current position (CP strategy) or waypoint (after time
  $dt$, TDP strategy). Unless otherwise specified, the $N-1$ chasers
  are evenly divided into four groups from $1$ to $4$ with their $t$'s
  in the corresponding sub-domains labeled accordingly if the
  cross-like domain is used.}
\end{figure}

Lastly, the tracking force of chaser $i$ contains two terms related to
the velocity and position of the target, respectively,
\begin{equation} \label{tracking_force}
\mathord{\buildrel{\lower3pt\hbox{$\scriptscriptstyle\rightharpoonup$}}
  \over F} _i^{track} = \alpha \hat v^t\Theta (l-r_i^{ct}) + \beta
\hat n_i^{ct},
\end{equation}
where the velocity-related term puts a force of magnitude $\alpha$
along the target's velocity unit vector $\hat v^t$ on chaser $i$ so
that it can follow the target if the distance $r_i^{ct}$ between them
is shorter than $l$. The position-related term is defined in
Fig. \ref{fig:algorithm}, where chaser $i$ is subject to a force of
magnitude $\beta$ along a unit vector $\hat n_i^{ct}$ pointing from
chaser $i$ at position ($x_i^c,y_i^c$) to a random target point
($x^t,y^t$) generated in a square domain of size $L$
(Fig. \ref{fig:algorithm}a) or a cross-like domain composed of four
identical sub-domains of size $L_2-L_1$ by $L_3$ , away from the
target by a minimal distance $L_1$ and oriented perpendicularly to one
another (Fig. \ref{fig:algorithm}b), centered on point $T$. Point $T$
is the current position ($x^T,y^T$) of the target or the waypoint
($x^T+v_x^Tdt,y^T+v_y^Tdt$) of the target moving with velocity $v^T$
after time $dt$, the time step for integrating
Eqn. \ref{newton_law}. In the field of biology \cite{hansen23}, the
former and the latter are called the classical pursuit (CP) strategy
and the target direction pursuit (TDP) strategy, respectively. When
using the cross-like domain, unless otherwise specified, we evenly
divide the $N-1$ chasers into four groups from $1$ to $4$ with each
group's $t$'s in the corresponding sub-domain labeled accordingly, and
the distance $L_1$ serves as a control variable. The maximum value of
effective $L_1$, beyond which a static target cannot feel the
existence of a static chaser, is $(d^t + d^c)/2 = 0.625d^t$. We set
$\alpha=1$, $l=0.3$, $\beta=10$, $L=20d^t$, $L_2=1.125d^t$, and
$L_3=0.75d^t$, unless otherwise specified.

The DEM simulations use the repulsive interaction diameter $d^t$ and
amplitude $\epsilon^t$ of the target and mass $m$ of a particle as the
reference length, energy, and mass scales, respectively. Without loss
of generality, we set $N=256$ in this study to reveal sufficient
dynamical details of the capturing process. The main conclusions do
not change if we use a minimal value of $N=5$, which allows one target
and four chasers.

\section{Results and Discussions}
\label{results_and_discussions}
We first test the target-chaser trapping system using the tracking
force whose TDP position-related term is associated with a square
domain and a single chaser group. Next, we replace the square domain
by a cross-like domain of varied geometry and the chasers are divided
into four groups. Finally, we investigate the effect of the tracking
force by turning off its velocity-related term or using the CP
position-related term.

\subsection{The target pursued by a single chaser group}
\label{single_group_chaser}
\begin{figure}
\includegraphics[width=0.40\textwidth]{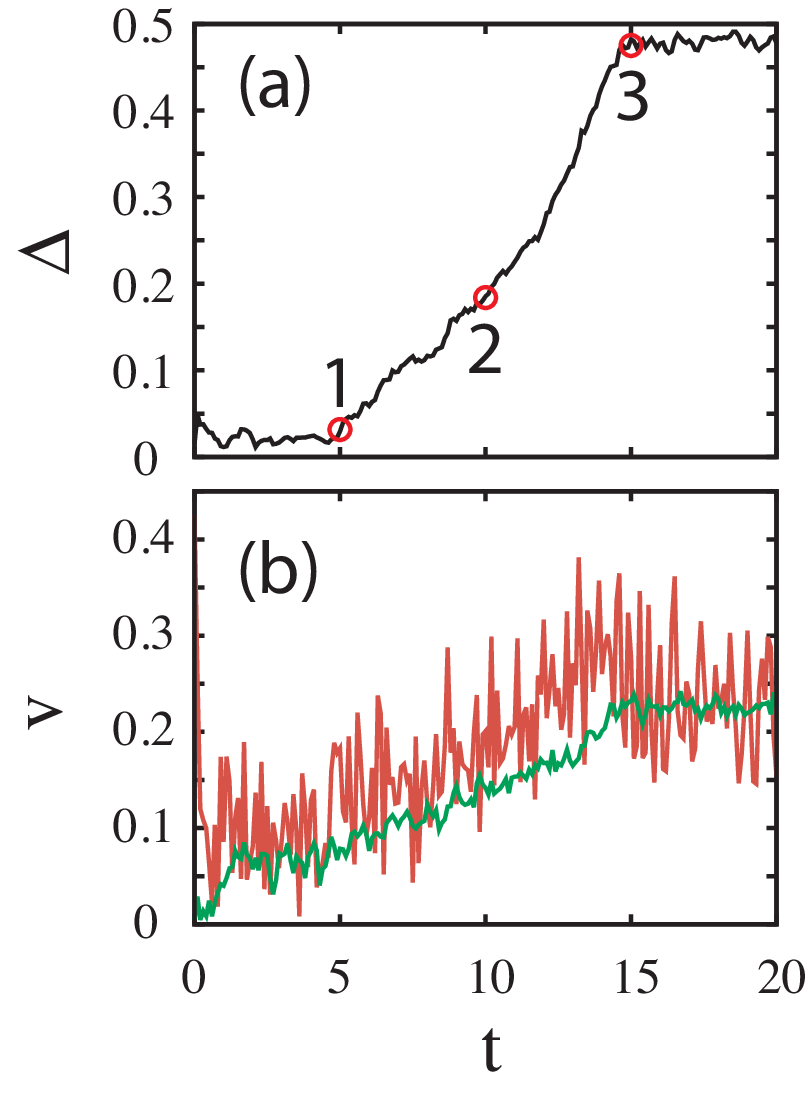}
\caption{\label{fig:1division} (Color online) Representative data of a
  failed capture. (a) Separation distance $\Delta$ between the target
  and the center of mass of a single chaser group as a function of
  time $t$. The corresponding configuration of the system when the
  value of $t$ is at 1, 2, or 3 (red open circle) is shown in
  Fig. \ref{fig:1division_details}. (b) Velocity $v$ of the target
  (red) or the center of mass of its chasers (green) as a function of
  time $t$. The data are obtained using a single realization.}
\end{figure}

To see if an escaping target can be trapped by a single chaser group,
we implemented the TDP position-related tracking force associated with
a square domain defined in Fig. \ref{fig:algorithm}a. The separation
distance $\Delta$ between the target and the center of mass of its
chasers as a function of time $t$ of a representative realization is
shown in Fig. \ref{fig:1division}a. The value of $\Delta$ is close to
zero for a short while at the initial stage, meaning the target is
well surrounded by its chasers. However, after that, $\Delta$
increases rapidly until it reaches a final steady value, where the
target, followed by the chasers behind, escapes. The velocity of the
target is always faster than that of the chaser group until reaching
the final stage, as shown in Fig. \ref{fig:1division}b. As a remark,
the CP tracking force leads to a similar but slower escape of the
target because chasers do not go to the waypoint of the target and
therefore the pursue is not as aggressive as in the TDP strategy.

\begin{figure}
\includegraphics[width=0.35\textwidth]{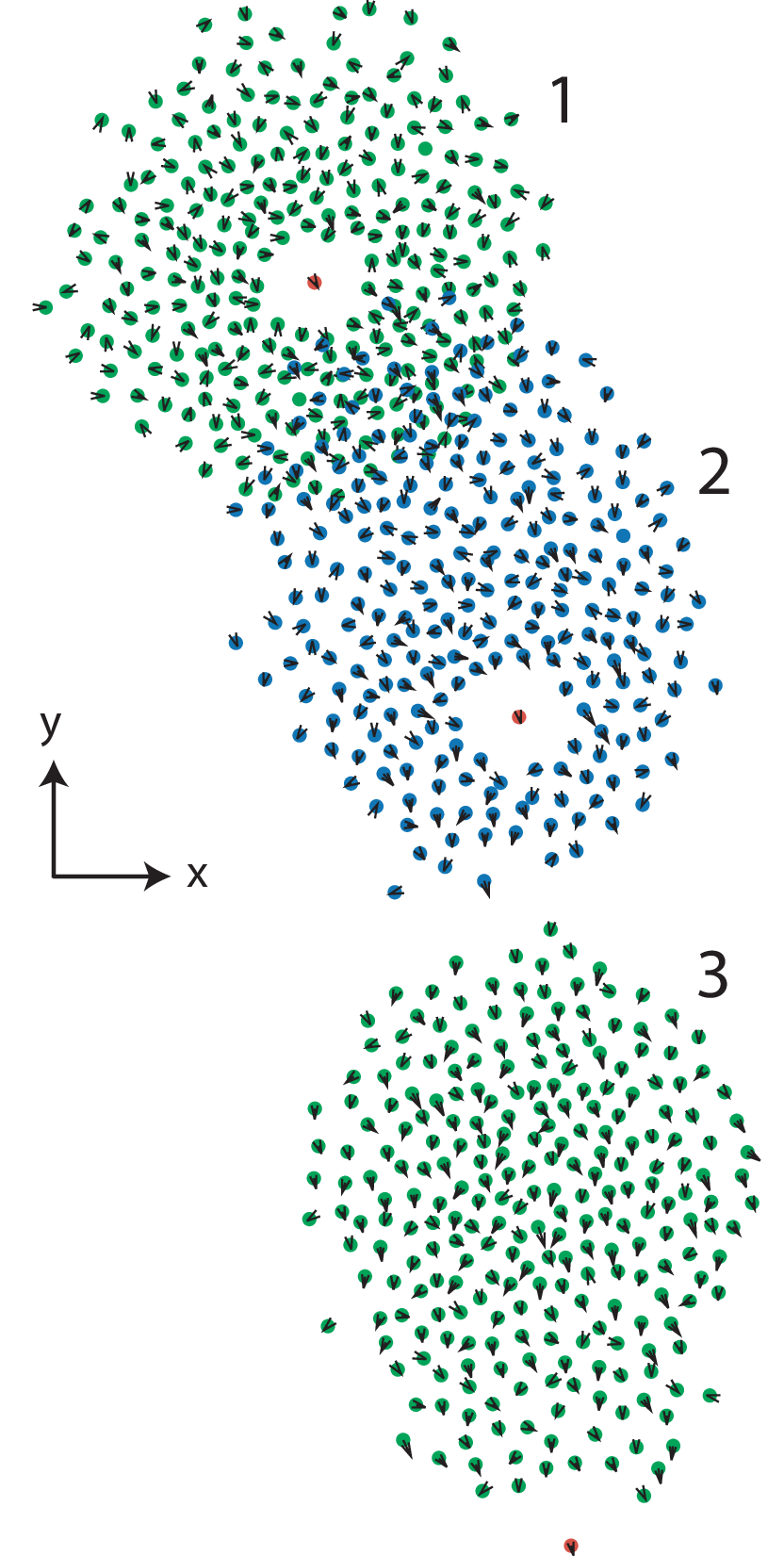}
\caption{\label{fig:1division_details} (Color online) Representative
  configurations of a failed capture, where the system is composed of
  one target (red) and a single chaser group (green or blue so that
  different configurations are visually discernible), when the value
  of time $t$ is at 1, 2, or 3, as labelled in
  Fig. \ref{fig:1division}a. All particles are drawn with identical
  arbitrary size and their relative positions preserved; added arrows
  show each particle's current direction of movement, with length
  being proportional to its speed.}
\end{figure}

In Fig. \ref{fig:1division_details}, we show the system at its
initial, middle, and final stages of escape, corresponding to time $t$
at 1, 2, and 3 labelled in Fig. \ref{fig:1division}a. The initial
stage (state 1) is unstable and therefore transient. The unstable
initial stage soon moves to the middle stage due to a force imbalance
between the target and chasers (state 2). Built-up force imbalances
speed up the separation process with positive feedback until the
target and the chasers reach a well-separated final stage (state 3),
where the target is always ahead of its chasers, and the whole system
becomes stable again.

\subsection{The target pursued by four chaser groups}
\label{multi_group_chaser}
\begin{figure}
\includegraphics[width=0.40\textwidth]{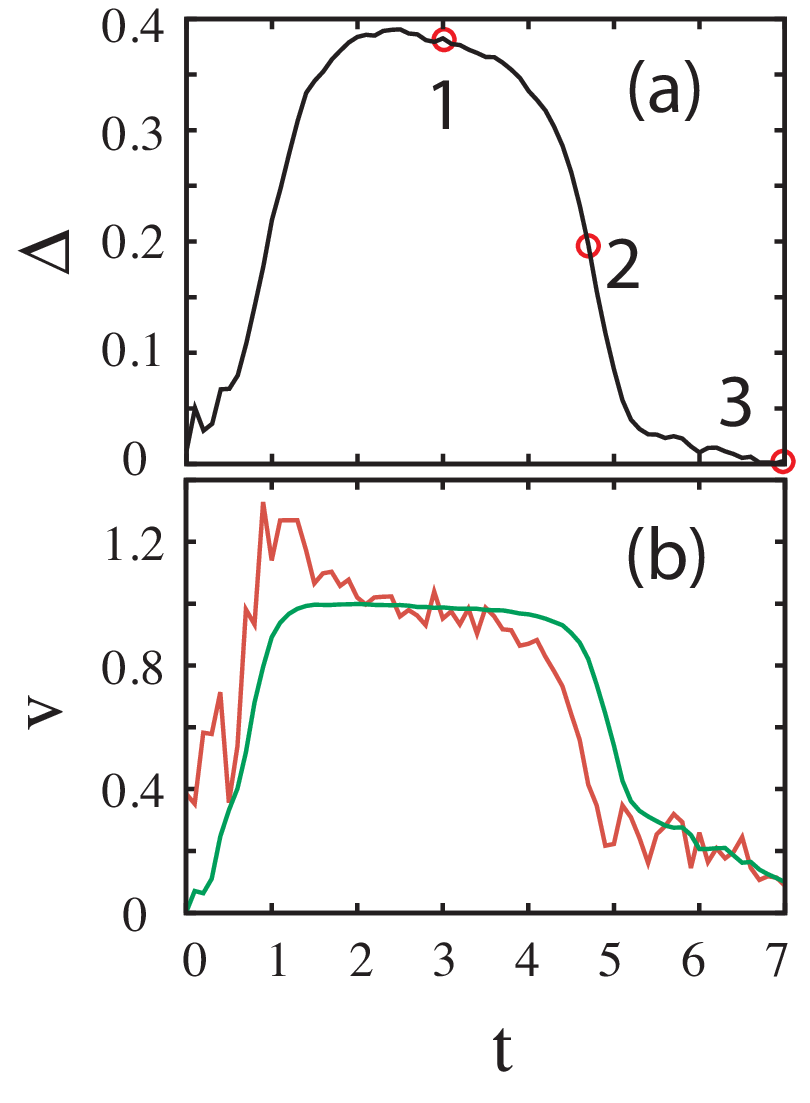}
\caption{\label{fig:4division_bis} (Color online) Representative data
  of a successful capture using the TDP position-related tracking
  force associated with a cross-like domain with $L_1=0.375d^t$ and
  four chaser groups. (a) Separation distance $\Delta$ between the
  target and the center of mass of the four chaser groups as a
  function of time $t$. The corresponding configuration of the system
  when the value of $t$ is at 1, 2, or 3 (red open circle) is shown in
  Fig. \ref{fig:4division_bis_details}. (b) Velocity $v$ of the target
  (red) or the center of mass of its chasers (green) as a function of
  time $t$. The data are obtained using a single realization.}
\end{figure}

Since an escaping target cannot be trapped by a single chaser group,
we then tried the TDP position-related tracking force associated with
a cross-like domain with $L_1=0.375d^t$ and four chaser groups,
defined in Fig. \ref{fig:algorithm}b. The separation distance $\Delta$
between the target and the center of mass of its chasers as a function
of time $t$ of a representative realization is shown in
Fig. \ref{fig:4division_bis}a. Unlike the failure of trapping the
target by a single chaser group, the four chaser groups form a stable
energy minimal of the repulsive potential, whose force form is defined
in Eqn. \ref{interparticle_rep_force}, and successfully capture the
target. The velocity of the target is initially faster than that of
the chasers but eventually the chasers catch up, slow down the target,
and arrest it, as shown in Fig. \ref{fig:4division_bis}b.

\begin{figure}
\includegraphics[width=0.35\textwidth]{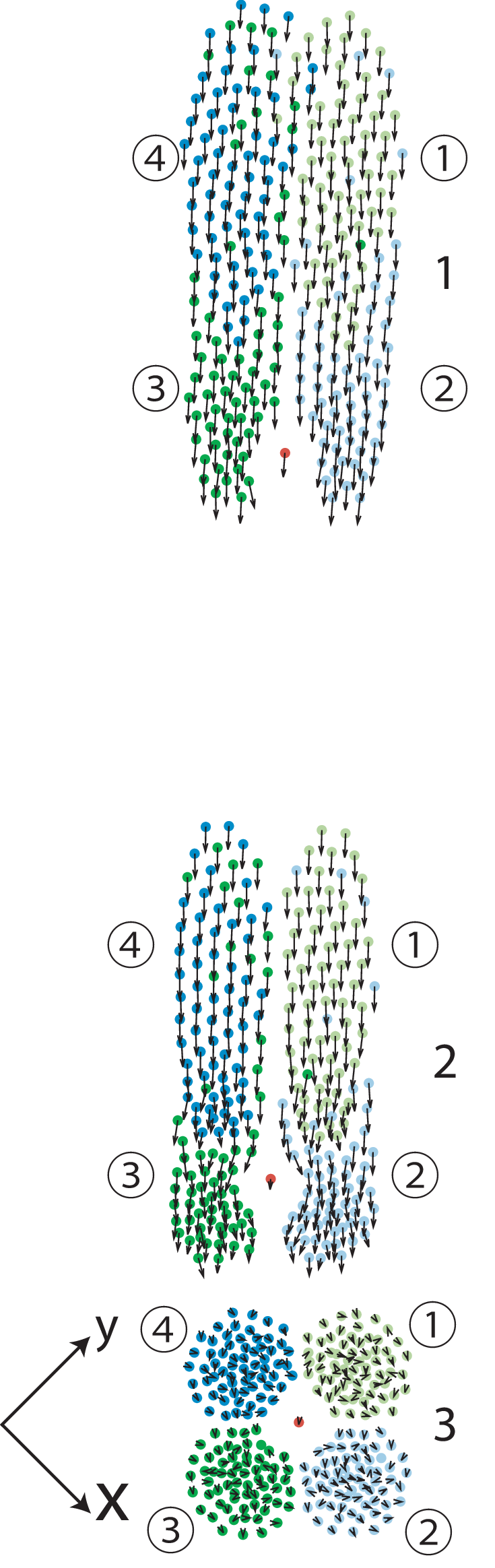}
\caption{\label{fig:4division_bis_details} (Color online)
  Representative configurations of a successful capture, where the
  system is composed of one target (red) and four chaser groups with
  $L_1=0.375d^t$ (1: light green, 2: light blue, 3: green, and 4:
  blue) when the value of time $t$ is at 1, 2, or 3, as labelled in
  Fig. \ref{fig:4division_bis}a. Except the additional group colors,
  all particles are drawn in the same way as in
  Fig. \ref{fig:1division_details}.}
\end{figure}

The capture mechanism of the four chaser groups can be explained using
Fig. \ref{fig:4division_bis_details}, which shows the system at its
initial, middle, and final stages of capturing, corresponding to time
$t$ at 1, 2, and 3 labelled in Fig. \ref{fig:4division_bis}a. At the
initial stage of capturing (state 1), the two chaser groups ($1$ and
$4$) at the tail of the target-chaser system push not only the target
but also the two other chaser groups (2 and 3) ahead of them. The
position-related tracking force $\beta \hat n_i^{ct}$, defined in
Eqn. \ref{tracking_force}, helps the two leading chaser groups (2 and
3) pass the target and converge in front of it (state 2). Once the two
leading chaser groups are ahead of the target, they are pushed and
accelerated not only by the two chaser groups behind but also by the
target. The leading chasers close to the equilibrium average positions
rapidly slow down and eventually the capturing process completes,
where the target stably sits in the potential energy minimum formed by
the four groups containing chasers equal in number and located
isotropically in its vicinity and cannot escape anymore (state 3).

\begin{figure}
\includegraphics[width=0.40\textwidth]{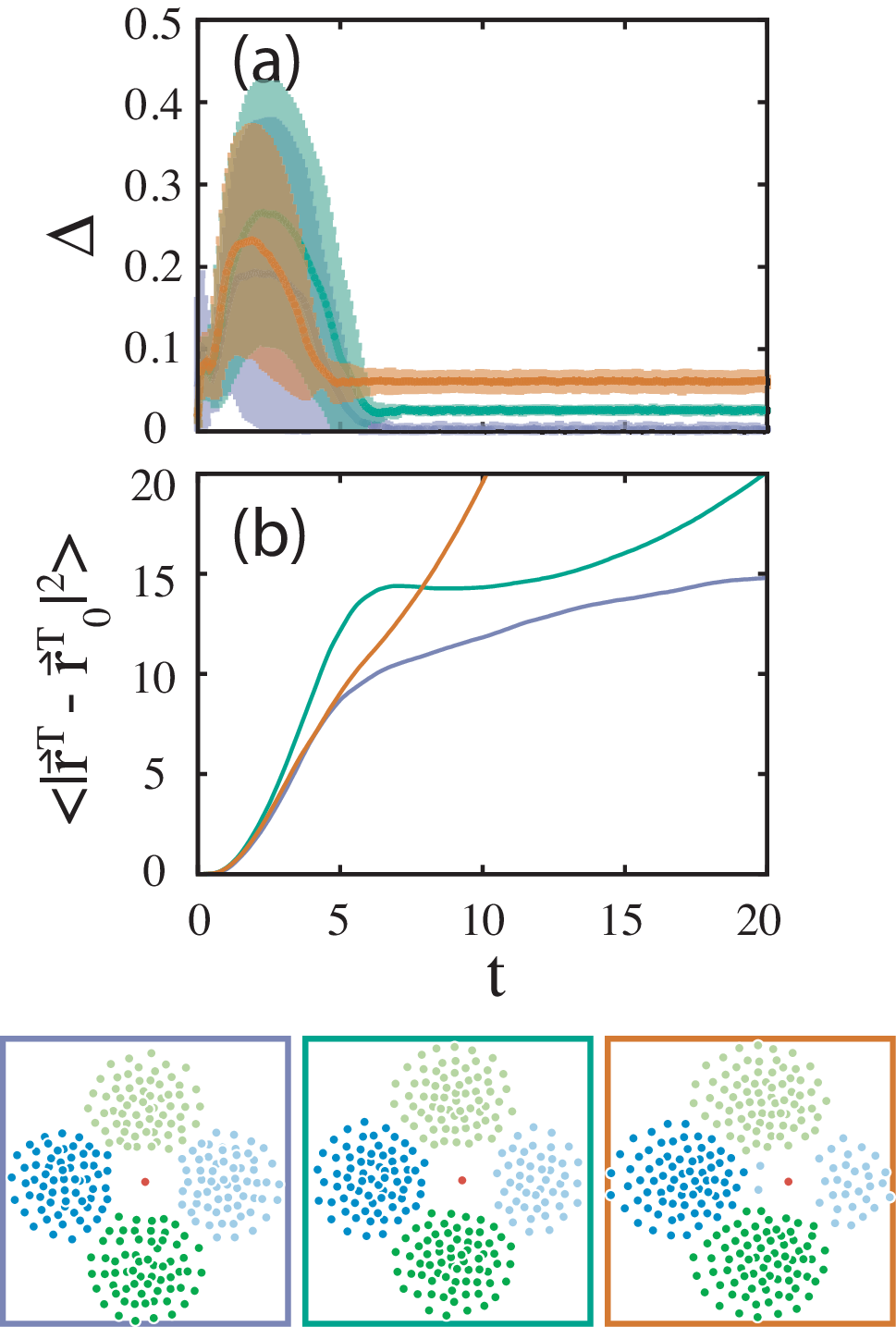}
\caption{\label{fig:4division_bis_varied_dist} (Color online) (a)
  Separation distance $\Delta$ between the target and the center of
  mass of four chaser groups with $L_1=0.375d^t$ as a function of time
  $t$. The number of chasers in one group (light blue in (b) below) is
  $63$ (purple), $47$ (green), and $31$ (orange) with the remaining
  chasers evenly divided into the other three groups. (b) The
  corresponding mean square displacement (MSD) of the target as a
  function of time $t$. The snapshots below, where particles are
  colored identically as in Fig. \ref{fig:4division_bis_details}
  within boxes of the same size and identical color code as in (a)
  acting as a guide to the eye, show the state of the system at
  $t=20$. Each mean, its error bar, and MSD are obtained using 10
  realizations.}
\end{figure}

To show that the equality in number of the four chaser groups is
indeed crucial for the stability of the system and a successful
capture, we increase the nonuniformity of the system by decreasing the
number of chasers in one group and evenly dividing the remaining
chasers into the other three groups. As a function of time $t$, we
plot the separation distance $\Delta$ between the target and the
center of mass of its chasers and the mean square displacement (MSD)
of the target,
$<|{{\mathord{\buildrel{\lower3pt\hbox{$\scriptscriptstyle\rightharpoonup$}}
      \over r}
  }^T}-{\mathord{\buildrel{\lower3pt\hbox{$\scriptscriptstyle\rightharpoonup$}}
    \over r} _0^T}|^2>$, where
${{\mathord{\buildrel{\lower3pt\hbox{$\scriptscriptstyle\rightharpoonup$}}
      \over r} }^T} = ({x^T},{y^T})$ and
${\mathord{\buildrel{\lower3pt\hbox{$\scriptscriptstyle\rightharpoonup$}}
    \over r} _0^T}$ are the current and the initial positions of the
target, respectively. With increasing inequality in number, the
separation distance $\Delta$ increases as well, as shown in
Fig. \ref{fig:4division_bis_varied_dist}a. Meanwhile, the MSD of the
target diverges with increasing nonuniformity because the target is
pushed by the unbalanced net repulsive force created by the nonuniform
chaser groups and the system never reaches a stable state, as shown in
Fig. \ref{fig:4division_bis_varied_dist}b.

\begin{figure}
\includegraphics[width=0.40\textwidth]{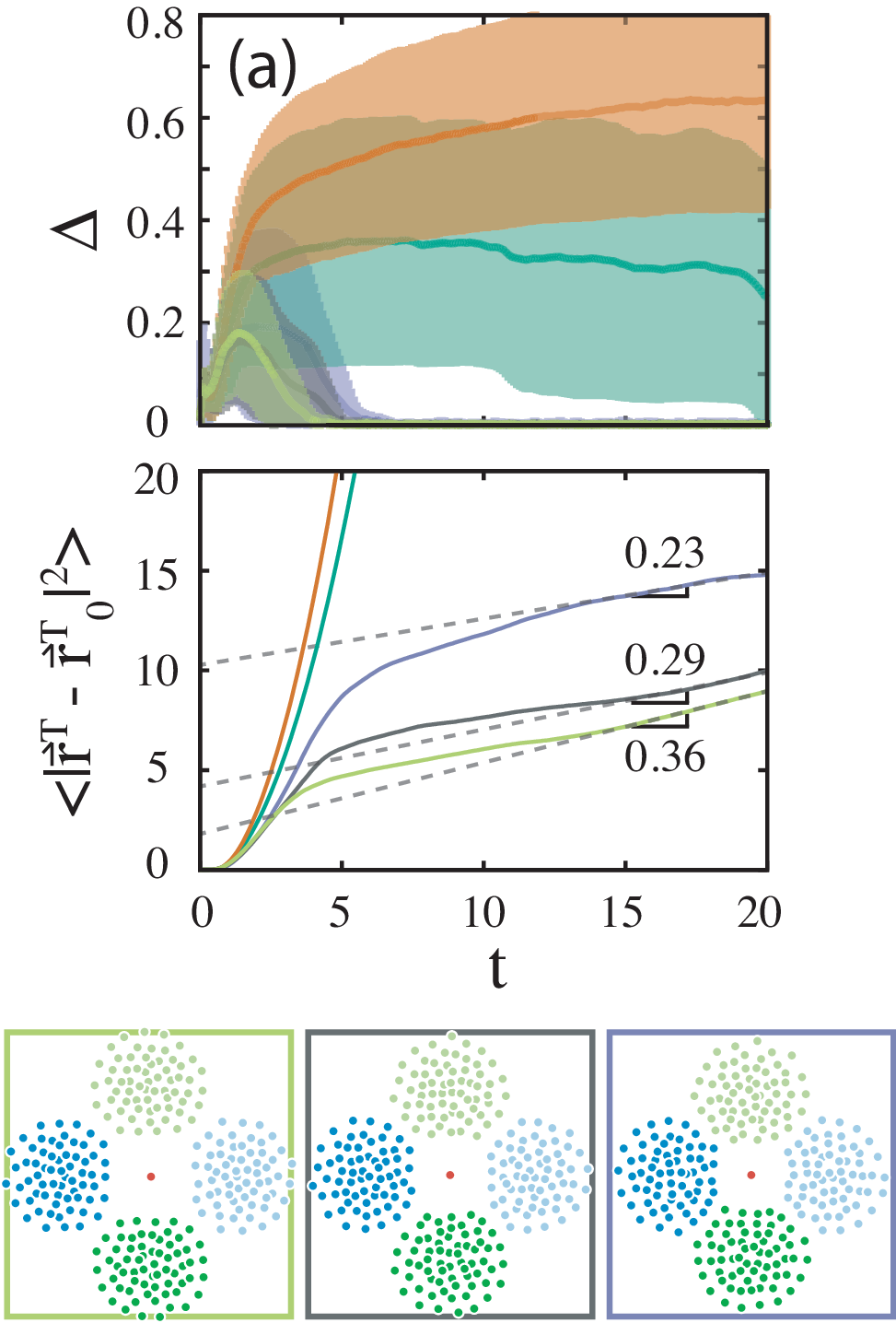}
\caption{\label{fig:separation_varied_limits} (Color online) (a)
  Separation distance $\Delta$ between the target and the center of
  mass of four chaser groups with $L_1=0.6255d^t$ (yellow green),
  $0.495d^t$ (grey), $0.375d^t$ (purple), $0.225d^t$ (green), and
  $0.150d^t$ (orange) as a function of time $t$. (b) The corresponding
  mean square displacement (MSD) of the target as a function of time
  $t$. Each dashed line with its slope shown is a linear fit of an MSD
  curve between $t=15$ and $20$. The snapshots below, where particles
  are colored identically as in Fig. \ref{fig:4division_bis_details}
  within boxes of the same size and identical color code as in (a)
  acting as a guide to the eye, show the state of the system at
  $t=20$. Each mean, its error bar, and MSD are obtained using 10
  realizations.}
\end{figure}

After showing that using four chaser groups can successfully capture
an escaping target, we investigate the effect of varying $L_1$, the
shortest distance measured from the cross-like domain to the target
for the position-related tracking force. Using smaller $L_1$ allows
chasers to get closer to the target and on average results in greater
target-chaser repulsive forces, which more likely push the target away
and slow down the capturing process. If the value of $L_1=0.375d^t$ is
reduced by a factor of $0.6$, the time required to capture the target
is increased by a factor of more than four. If the value of $L_1$ is
further reduced by a factor of $0.4$, a successful capture is
impossible. The results are shown in
Fig. \ref{fig:separation_varied_limits}a. To learn if there exists a
minimum $L_1$, where a stably trapped target is least able to move, we
plot the corresponding MSD as a function of $L_1$. After the system
reaches its steady state, the slope of the MSD, proportional to the
diffusion constant, decreases from $0.36$ to $0.23$ as $L_1$ decreases
from $0.6255d^t$ to $0.375d^t$ and then diverges with further decrease
of $L_1$. The results, shown in
Fig. \ref{fig:separation_varied_limits}b, indicate that forming a
tighter trap, within which the random walk step size of the target is
smaller, makes it less mobile and an optimal $L_1$ exists.

\begin{figure}
\includegraphics[width=0.40\textwidth]{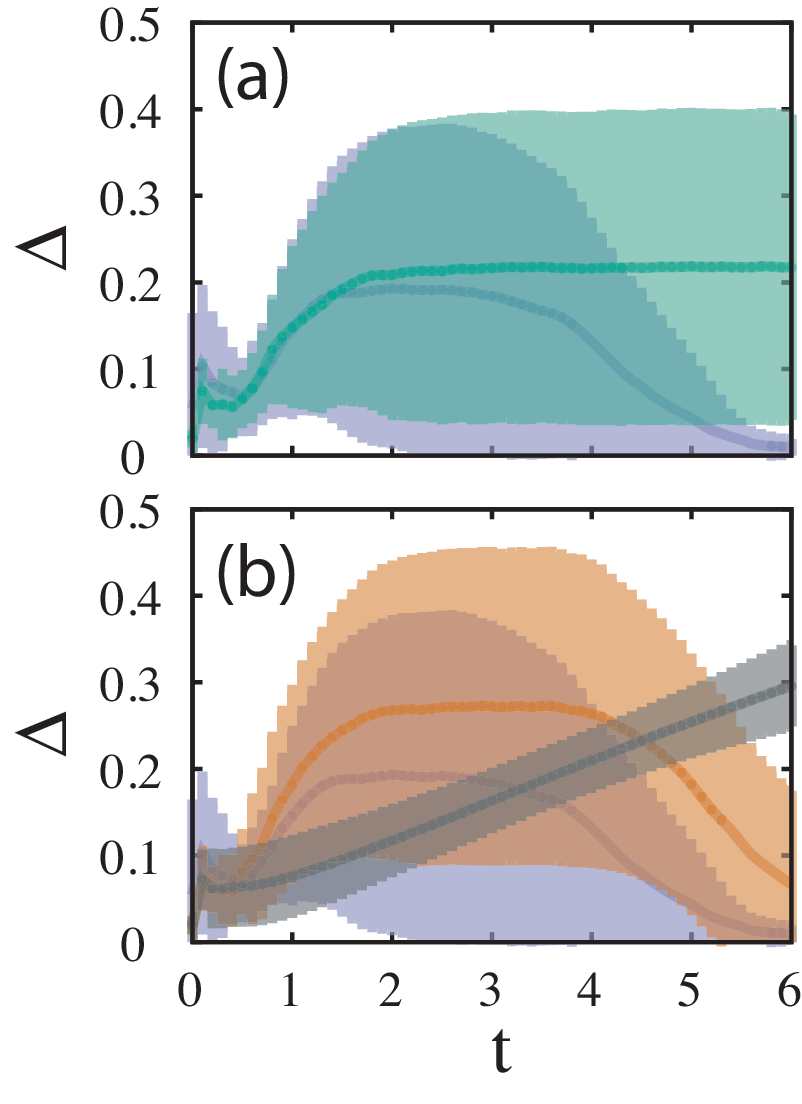}
\caption{\label{fig:separation_varied_types} (Color online) (a)
  Separation distance $\Delta$ between the target and the center of
  mass of the four chaser groups with $L_1=0.375d^t$ subject to the
  position-related TDP tracking force ($\beta=10$) without
  ($\alpha=0$, green) or with ($\alpha=1$, purple) the
  velocity-related tracking force as a function of time $t$. (b)
  Separation distance $\Delta$ between the target and the center of
  mass of the four chaser groups with $L_1=0.375d^t$ subject to the
  velocity-related tracking force ($\alpha=1$) without ($\beta=0$,
  grey) or with ($\beta=10$) the position-related TDP (purple) or CP
  (orange) tracking force as a function of time $t$. Each mean and its
  error bar are obtained using 10 realizations.}
\end{figure}

To learn the importance of the velocity-related term and the
position-related term in the tracking force, we change the value of
$\alpha$ and $\beta$ in Eqn. \ref{tracking_force}
independently. First, switching off the velocity-related term in the
tracking force by changing the value of $\alpha$ from $1$ to $0$ makes
a successful capture impossible, as shown in
Fig. \ref{fig:separation_varied_types}a. Then we change the the value
of $\beta$ and the type of position $T$. Switching off the
position-related term in the tracking force by changing the value of
$\beta$ from $10$ to $0$ also makes a successful capture impossible,
as shown in Fig. \ref{fig:separation_varied_types}b. Additionally,
changing the type of position $T$ from TDP to CP strategy damages the
efficiency of capture, because the CP strategy does not allow chasers
to predict where the target will be at the next moment. Overall, both
the velocity-related term and the position-related term in the
tracking force play a crucial role in the capturing process because it
allows chasers to follow the target closely. Besides, guiding chasers
to the target's future position by using the TDP position-related term
in the tracking force improves the efficiency of capture, which is
desirable but not essential.

\vspace{6 mm}
\section{Conclusions}
\label{conclusions}
In this study, we propose a dynamical trap made of chasers subject to
target-tracking forces using the velocity and position information of
the target. The velocity-related tracking force allows a chaser to
synchronize its moving direction with that of the target while the
position-related tracking force guides a chaser to the vicinity of the
target. We show that the strategy of dividing the chasers into four
groups and assigning each group to a designated domain close but not
immediately next to the target allows successful and inescapable
capture. On the other hand, using a single chaser group creates
unbalanced forces between the target and its chasers and therefore the
capturing process becomes unstable and fails. Furthermore, we also
show that the velocity-related tracking force is also crucial to the
dynamical trap, while the position-related tracking force able to
guide chasers to the future instead of the current position of the
target only improves the efficiency of the capturing process.

Potential applications of our study include capturing wild animals
such as hungry bears entering areas inhabited by humans using unmanned
aerial vehicle (UAV) like drones \cite{souza22,sarmento25}. Multiple
chaser drones can be controlled by a commanding one that monitors
their positions and velocities using reconstructed three-dimensional
images, a technique for studying the group behavior of birds
\cite{ballerini08}. Besides, a target animal can be tracked using
drone-mounted thermal infrared cameras \cite{burke19}. To realize
these applications, it is essential to study how a bear responses to
multiple approaching chasers and if there exists an uncatchable
situation even though in principle the trapping system is unbreakable,
which will be the future direction of our work.

\section{acknowledgments}
GJG thanks Shizuoka University for supporting this work and the
reviewer for insightful comments.

\bibliography{71812}

\end{document}